\documentstyle[aaspp4]{article}
\received{00 March 1996}
\accepted{00 1996}
\def\ee #1 {\times 10^{#1}}          
\def\ut #1 #2 { \, \mathrm{#1}^{#2}} 
\def\u #1 { \, \mathrm{#1}}          

\def\kms{km s$^{-1}$}

\leftline{To appear in {\it The Astrophysical Journal}\hfill}

\righthead{Radio and Near-IR  Observations of The Trifid Nebula}

\begin{document}

\title{Radio  Continuum Emission from the \\
Central stars of M20 and the Detection  \\
of a New  Supernova Remnant Near M20}

\author{F. Yusef-Zadeh}
\affil{Department of Physics and Astronomy, Northwestern University, 
Evanston, Il. 60208 (zadeh@nwu.edu)}

\author{Mark Shure\footnote{Visiting Astronomer, NASA Infrared Telescope 
Facility, which is operated
by the University of Hawaii under contract with the National Aeronautics
and Space Administration.}}
\affil{Center for High Angular Resolution Astronomy,
Georgia State University, Atlanta GA  30303 (shure@chara.gsu.edu)}

\author{Mark Wardle}
\affil{Research Centre for Theoretical Astrophysics, University of Sydney, 
Sydney, NSW 2006, Australia (wardle@physics.usyd.edu.au)}

\author{N. Kassim}
\affil{Naval Research Lab., Washington, DC 20375-5351 
(nkassim@shimmer.nrl.navy.mil)}

\begin{abstract}

The Trifid nebula (M20) is a well-known prominent optical HII region 
trisected by obscuring dust lanes.  Radio continuum VLA observations 
of this Nebula show free-free emission at $\lambda$3.6 and 6cm from 
three stellar sources lying close to the O7V star at the center of the 
nebula.  We argue that neutral material associated with these stars 
are photoionized externally by the UV radiation from the hot central 
star.  We also report the discovery of a barrel-shaped SNR G7.06--0.12 
at the northwest rim of the nebula and two shell-like features 
G6.67--0.42 and G6.83--0.21 adjacent to W28 and M20.  We discuss the 
nature of these features and their possible relationship to the pulsar 
PSR 1801-2306 and W28 OH (1720 MHz) masers.

\end{abstract}

\keywords{galaxies:  ISM---Galaxy:  ---ISM: individual 
(the Trifid Nebula) --- ISM: HII regions}

\vfill\eject

\section{Introduction}

The Trifid Nebula (M20), one of the most spectacular optical HII 
regions in the sky, is centered on a small cluster of hot stars which 
include components A through G of HD 164492.  M20 is located at a 
distance of $\approx$1.7 kpc in the Sagittarius spiral arm (but see 
also Kohoutek, Mayer \& Lorenz 1999); it's angular size of 6$'$ 
corresponds to about 3pc at this distance.  The ionizing flux of 
$10^{48.8}$ s$^{-1}$ required to maintain the HII region (Chaisson and 
Wilson 1975) is supplied by the O7.5III star HD 164492A (Wallborn 
1973), which has M$_v$=--5.3 for A$_v\approx$1.3 towards the central 
stars (Lynds and O'Neil 1985).  This nebula is associated with the 
young star cluster NGC 6514 and a molecular cloud to the southwest 
(Ogura and Ishida 1975).  The recent detection of several molecular 
condensations associated with protostellar sources in the HII region 
and HH 399, a remarkable jet-like structure, suggest a new generation 
of star formation induced by the nebula (Cernicharo et al.\ 1998).  
Hester et al.\ (1999) have recently shown the high resolution images 
of the SE corner of the nebula based on observations made with WFPC2 
of the HST. The emission from this corner of the HII region is 
dominated by a photoionized photoevaporative flow.

M20 shows many similarities to M42 such as its interaction with its 
parent molecular cloud and its inhomogeneous nebular structure, but 
the Trifid is thought to be significantly younger than the Orion 
Nebula ($\sim 10^{5}$ yr rather than $\sim 10^{6}$ yr) and the 
protostellar molecular condensations associated with massive star 
formation are even younger (10$^4$ yr) (Cernicharo et al.\ 1998).

Here we present near IR and radio observations of M20.  We have 
detected a number of near-IR stellar sources within the central star 
cluster in J, H, K and L images.  All seven of the components of HD 
164492 (A--G) appear to varying degrees in these images.  Three 
components -- B, C and D -- coincide with compact radio continuum 
sources within 12 arcseconds of HD 164492A. These detections suggest 
that the central massive star is photoionizing the envelope of cool 
stars in its immediate vicinity similar to the ``proplyds'' in the 
Orion Nebula (Churchwell et al.\ 1987; Garay et al.  1987; O'Dell et 
al.  1993).  Radio images of the nebula at 20 and 6 cm show dark 
features suggesting the presence of cold and dense regions of dust and 
gas clouds within the HII region shadowing the UV radiation from HD 
164492A. Lastly, on a scale of tens of arcminutes, we report the 
discovery of a new candiate barrel-shaped supernova remnant (SNR) 
lying adjacent to M20 and two shell-type features to the north and east of 
SNR W28.

\section{Observations}

The Very Large Array of the National Radio Astronomy 
Observatory\footnote{The National Radio Astronomy Observatory is a 
facility of the National Science Foundation, operated under a 
cooperative agreement by Associated Universities, Inc.} was used in 
its compact C and D configurations at $\lambda$20, 6 and 2cm in 1987 
(Yusef-Zadeh et al.  1991).  Follow-up high resolution observations in 
the wide BnA array were also carried out at 3.6, 6 and 20cm in 1998.  
In all observations 3C286 was used as standard amplitude calibrator 
and 1748-253 and 1730-130 as phase calibrators.  The 2 and 6cm data 
were successfully self-calibrated in the BnA array configuration.  The 
$\lambda$20cm and 6cm data corresponding to each configuration were 
combined before the final images were constructed using VTESS in AIPS. 
In order to measure the polarization characteristics of G7.06--0.12,
 we 
used a different antenna pointing at 6cm centered at $\alpha, 
\delta(1950) = 17^h 59^m 18^s, -22^0 55' 45''$ which is offset to the 
north of the central star HD 164492A.

Wide-field $\lambda$90cm observations towards the supernova 
remnant W28 which includes M20 were also obtained in the D
configuration of the VLA in 1996. In this case 3C48 and 3C286
were used as amplitude calibrators, while 1828+487 (3C380) was
used for phase calibration. The 90 cm data were processed using
the wide-field imaging task DRAGON present in the NRAO Software
Development Environment (SDE) reduction package. DRAGON
successfuly completed 5 loops of self-calibration and CLEAN-based
deconvolution to generate the final $\lambda$90cm image.

J (1.2 $\mu$m), H (1.6 $\mu$m), K (2.2 $\mu$m) and L (3.5 $\mu$m) 
images were obtained in July 1995 using the near-IR array camera 
NSFCAM at the NASA Infrared Telescope 
Facility\footnote{The NASA Infrared Telescope Facility is operated under 
contract by the University of Hawaii.}.  Due to the short time spent 
taking these images, some of the brighter stars were slightly 
saturated and stars fill the blank sky images.  However, the main 
results presented here are not affected by the quality of the images.


\section{Results and Discussion}

\subsection{Central Stars of M20}

Figure 1 shows contours of 3.6cm emission with a resolution of 
0.63$''\times0.45''$ (PA=80$^0$) with rms$\approx33 \mu$Jy.  Radio 
continuum peaks whose flux densities and sizes are listed in Table 1 
coincide with the positions of HD164492 B, C and D. Optical positions 
of HD164492 A, B and C are also listed in Table 1 using the Hipparcos 
coordinates which are based on a fit to a multiple source model.  The 
quality of the model solution for "C" using the Hipparcos database was 
poor, so we did not include it.  However, there is an excellent 
agreement between the absolute position of radio and optical sources 
to within 0.1$''$, especially for sources B and C.
  
The grayscale images in Figure 2 show the components of multiple star 
HD 164492 and are identified on the J image, which is very similar to 
visible images.  The central star A coincides with the optical star HD 
164992A but is not detected at a level of rms$\approx33 \mu$Jy at 
3.6cm.  Source C, the brightest radio source, has a flux of 2.43 mJy 
at 3.6cm.  HD 164492D lies only 2 arcsec west of star C and was found 
to be a strong H$\alpha$ emission source by Herbig, who classified it 
as a Be-star and named it LkH$\alpha$ 123 (Herbig 1957).  It was also 
included as H$\alpha$-emission star number 46 in a survey by Velge 
(1957) and star number 145 in the survey of Ogura and Ishida (1975).  
The D star is the brightest point source in recent 12.5 and 17.9 
$\mu$m images that were taken with the JPL MIRLIN camera at the IRTF 
(Ressler and Shure 1995).  Notice several objects in the L image which 
are either unseen or are very much dimmer shortward of K. Among these 
newly discovered sources is one roughly 3 arcsec NE and the other only 
2 arcsec N of star A. If they are physically associated with this star 
(2$''$ = 3400 AU at 1.7 kpc), they would represent some of the first 
low-mass companions to high-mass stars.

All three sources B, C and D are also detected at 6cm based on our 
high resolution BnA array data.  Accurate spectral index measurements 
between 6 and 3.6cm using similar {\it uv} coverage and spatial 
resolution of 0.99$''\times0.7''$ (PA=--87$^0$) showed 
$\alpha\approx$0 for sources B and C and $\alpha$=--0.19, where 
F$_\nu\propto\nu^\alpha$, for source D. The D star as listed in Table 
1 is the only star resolved in our 3.6cm measurements with a 
deconvolved size of 0.19$''\times0.12''$ (PA=129$^0$).

\subsection{Dense neutral gas associated with stars B,C, and D}

What is the origin of the emission from B,C, and D? By analogy with 
the Orion nebula, one suspects that these sources are externally 
ionised neutral condensations being photoevaporated by the intense UV 
field of the O7.5III star HD 164492 A. The argument that this is so 
proceeds similarly to those of Garay et al.\ (1987) Churchwell et al.\ 
(1987) applied to the proplyds first detected in radio continuum in 
the Orion nebula.

The flat spectra of sources B,C, and D indicate that their radio 
continuum arises from optically thin free-free emission.  For 
$T\approx8000$\,K, the volume emission measures are $n_e^2 V \sim 1 
\ee 57 \ut cm -3 $, and for the emission to be optically thin the 
characteristic scale $R$ of the emission region $\ga 50 \u AU $.  If 
this region is a roughly constant-density compact H\textsc{ii} region, 
this scale represents the outer boundary.  However, the gas is then 
gravitationally unbound to the central star and the H\textsc{ii} 
region would expand on a time scale of $\sim 30$ yr.  On the other 
hand, if the emitting region is an ionized wind with $n_e \propto 
r^{-2}$, the emission is dominated by the innermost radii and $R$ 
corresponds to the \emph{inner} boundary.  An ionized stellar wind can 
be discounted, because then $R\ll 50 \u AU $, and the mix of 
optically-thin and optically-thick contributions at any given 
frequency produces a $\nu^{0.6}$ spectrum; further the source would be 
much weaker than observed.  Thus we conclude that the emitting region 
is an ionized wind that is photoevaporated from a reservoir of neutral 
material near the star.

The neutral reservoir cannot be too large as sources B and C are 
unresolved and source D is barely resolved.  Adopting a distance of 
1.7 kpc to M20, the geometric mean of the semi-major and semi-minor 
axes of source D (see Table 1) is $R \approx 130 AU$, and the 
corresponding upper limit on sources B and C is 80 AU. The neutral 
reservoir will be even smaller, and is therefore clearly associated 
with the star.

The O7.5III star HD 164492 A is likely to be the dominant source of 
the ionizing photons.  An O7.5III star emits $9.6\ee 49 $ ionizing 
photons $\ut s -1 $ (Panagia 1973), so that the total ionizing flux 
incident on a 130 AU radius target at the projected distance of star D 
($\approx 0.095 \u pc $) is $1.1\ee 45 \ut s -1 $.  This is comparable 
to the hydrogen recombination rate of source D, $6.6\ee 44 \ut s -1 $.  
The recombinations in source B and C are also consistent with this 
hypothesis.  Note, however, that stars HD 164492 B,C, and D are of 
spectral type B (Kohoutek et al 1999) and produce $\sim 10^{45}$ Lyman 
continuum photons $\ut s -1 $ or more, so they may contribute 
significantly to the ionization if enough photons can intercept the 
nearby neutral material.

Our observations do not determine the distribution of the neutral material 
associated with stars B,C, and D, but we speculate that it is in 
circumstellar disks, as for the ``proplyds'' in Orion.

\subsection{Embedded EW Dust Lane in M20}

A $\lambda$6cm grayscale image with resolution of 2.3$''\times1.1''$ 
(PA=--80$^0$) and an optical image based on the Palomar Sky Survey are 
compared in Figures 3a and b.  The prominent elongated dust lanes to 
the SW, SE and NW of the optical image have no counterpart in radio, 
indicating that these dust features lie in front of the HII region.  
However, there is a remarkable EW dark radio feature which closely 
mimics the shape of the optical dust lane seen to the SW of the 
central hot star in Figure 3b near $\alpha, \delta(1950) = 17^h 59^m 
17^s, -23^0 02' 50''$.  Figure 4 shows a NS slice cut across this 
feature in the 6cm radio image.  The radio continuum emission is 
depressed by a factor of 3 where the optical nebula appears to be 
crossed by an EW dark dust lane with a thickness of about 40$''$ (0.3 
pc).  The correlation between reduced radio emissivity and the optical 
dust lane is evidence for the dust lane being embedded within the 
nebula.  Other examples of dark features are also apparent in the 
inhomogeneous large-scale distribution of ionized gas beyond the inner 
region shown in Figure 3.  These dark features are particularly 
noticeable as broken shell-like structures surrounding the 6$'$ size 
of the nebula and correlate with the distribution of the HCO$^+$ 
J=1--0 emission from the nebula presented by Cernicharo et al.  
(1998).  We also note a column of dark feature labelled as
dark shadow within the nebula to the east of TC1.
These dark features are
unlikely to be 
produced by a lack of short {\it uv} spacing data but are instead due 
to dense 10$^4$ cm$^{-3}$ column of gas arising from the surface layer 
of the molecular cloud and causing the HII region to become ionization 
bounded.  The dearth of emission from a series of dark features 
including the EW feature, as best represented in Figure 3a, are 
interpreted to be the peaks of dense gas shielding the ionizing flux 
arising from the central hot star.  These columns of dense gas are 
responsible to reduce the emission measure $n_e^2 L$ where $n_e$ and L 
are the electron density and the path length.

\subsection{Ionized Rims of TC1 and TC2}

Figure 5 shows total intensity contours over the central part of the 
HII region at 2.3$''\times1.1''$ resolution (PA$=-80^0$).  The two 
extended ionized features to the NW and SE are associated with two 
bright point-like condensations of dust emission at 1.3mm denoted TC1 
and TC2 by Cernicharo et al.\ (1998).  High velocity broad wings in 
the HCO$^+$ emission from TC1 and a jet-like HH feature associated 
with TC2 led Cernicharo et al.\ to suggest that these condensations 
are associated with protostars.  The age of these condensations are 
estimated to be about 10$^4$ yrs and therefore formed after the birth 
of the HII region.  The extended photoionized features in Figure 5 
delineate the ionized rims of TC1 and TC2 facing the central hot 
star.  High resolution WFPC2 observations of TC2 using a number of 
spectral lines was recently reported by Hester et al.  (1999) who 
interpret the ionized layer of TC2 as  a photoionizing 
photevaporative flow.  The typical flux density of the ionized rims of 
TC1 and TC2 is about 0.5 mJy/beam which corresponds to $n_e 
\approx2\times10^3$ cm$^{-3}$.

\section{Discovery  of a Supernova Remnant and two Shell-like Features}
\null

\subsection{G7.06--0.12: A Barrel-shaped SNR}

Figure 6 shows a large-scale grayscale image of M20 and its immediate 
vicinity at 20cm (D array only) with a resolution of 
$59.1''\times33.7''$ (PA=--2$^0$).  We note the prominent W28 SNR to 
the SW and report here a newly discovered shell-type SNR, G7.06-0.12, 
with a diameter of 13$'$ adjacent to the NW rim of M20.  Radio 
continuum measurements at 6cm show linearly polarized emission at the 
northern rim of M20 near $\alpha, \delta(1950) = 17^h 59^m 22^s, -22^0 
55'$.  The degree of polarization is about 28\% at 6cm based on D 
array data with a resolution of 29.8$''\times14.3''$ (PA=-27$^0$).  
This radio source has no obvious optical counterpart and is a 
previously uncataloged, synchrotron emitting SNR.

Further evidence that G7.06-0.12 is a SNR is provided by its 
morphology, revealed in the contour map of the 20cm image centered on 
the remnant and displayed in Figure 6a.  Here G7.06--0.12 
is revealed as 
a classic barrel shaped SNR aligned roughly with its major axis 
parallel to the Galactic equator (Gaensler 1998).  The western rim of 
the barrel is more extensive and is about 3-4 times brighter than the 
eastern rim, which protrudes north of M20 near $\alpha= 17^h 59^m 
25^s$.  We also note several clumps of 20cm continuum features to the 
north of 7.1--0.1 and a 8$'$ elongated structure running north-south 
toward the center of the SNR.

Finally, the identification of G7.06--0.12 as a barrel-shaped SNR is 
confirmed by its nonthermal continuum spectrum and morphology as 
determined from complimentary 90cm observations.  Figure 7b is D 
array, 90cm contour image centered on the bright SNR W28 with an 
angular resolution of 5.5$'\times2.6'$(PA=13.4$^0$).  Protruding north 
of the W28 SNR near $\alpha= 17^h 58.5^m$ is the western half of 
G7.06--0.12, whose eastern half is now smothered by M20 centered near 
$\alpha= 17^h 59.3^m$.  G7.0.6--0.12's barrel shaped morphology is 
also revealed on reinspection of the higher resolution ($\sim20''$) 
wide-field 90cm image centered near the SNR G5.4-0.1 (Frail, Kassim, 
and Weiler 1994).  The 90cm image is dominated by W28, whose 
integrated flux is 350$\pm$50 Jy.

Confirmation of the nonthermal spectrum of G7.06--0.12 comes from 
comparing the integrated flux of the well defined western half of the 
barrel, giving an average of $\sim$3.9 Jy as obtained from the two 
90cm measurements and $\sim$1.9 Jy from the 20 cm image.  This yields 
the canonical $\alpha \sim$-0.5$\pm$0.15 nonthermal spectrum of a 
typical shell-type SNR, as expected.  The images are sufficient to 
confirm that the eastern half of G7.06--0.12 is also nonthermal, but 
higher resolution 90cm data is required to determine the spectrum of 
either side of the barrel more accurately than this.  It is 
interesting to note that the integrated flux of M20 from our D array 
90 cm map, which includes some flux from the eastern half of 
G7.06--0.12, is 11$\pm$2 Jy.  Together with the 14$\pm$3 Jy integrated 
flux present on our 20 cm map this indicates a slightly inverted 
spectrum revealing that the HII region has started to become optically 
thick at 90 cm as is common.

While the eastern rim of G7.06--0.12 merges with the northwest segment 
of M20, there is no morphological evidence for the interaction of the 
SNR and M20.  We cannot rely on $\Sigma-D$ relationship to estimate 
the distance to G7.06--0.12 (Green 1991) but VLA observations at 74 
MHz should be able to determine the relative positions, since at that 
frequency M20 would have become completely optically thick and it 
would be apparent whether the eastern side of the barrel was being 
absorbed or not.  We note this region of the Galaxy is rich in a 
having a number of HII complexes ( M8, M20), the SNR W28 expanding 
into an adjacent molecular cloud (e.g.  Wooten 1981) as well as a 
young star cluster NGC 6514.  Thus, it is plausible that SNR 
G7.06--0.12 is associated with the young cluster, but further studies 
are needed.

\subsection{G6.67-0.42: A Nonthermal Shell-type Feature}

An additional new radio continuum source identified in this rich 
region of the Galaxy is G6.67--0.42, located 20$'$ south of M20 as 
labelled in Figure 6.  G6.67--0.42 is clearly distinct from W28 and 
shows a protruding feature in the eastern part of W28 near 
$\delta(1950) = -23^0 23'$.  This feature which is curving convex 
downwards is another previously, unidentified source.  It's morphology 
is clearly shell-like, as indicated from the 20 cm image displayed in 
Figure 7a.  However the limitation of the 20 cm field of view together 
with the confusion from W28 allows only a northeastern fragment of an 
apparently circular shell to be clearly delineated.  High-resolution 
observations at 20cm reveals this new feature which otherwise would 
have been associated to W28 in earlier low-resolution measurements.  
The 90cm image shown in Figure 7b shows G6.67--0.42 merged with the 
eastern part of the shell in W28.

Estimating the spectrum of G6.67--0.42 was done by convolving the 20 
and 90cm images to the same resolution 328.8$''\times$ 157.6$''$.  The 
spectrum is estimated to be $\approx-$0.35$\pm$0.15 from the peak flux 
and integrated flux of the shell fragment at 20 and 90cm.  The 
nonthermal emission could be supplied or be part of the W28 remnant 
itself, such as resulting from a so-called "blow-out" where the main 
blast wave has encountered a lower density ISM region.  However, the 
geometry of the weaker shell-like feature and its distinct structure 
from W28 is not classic blow-out morphology in that its implied center 
of curvature is significantly different than that of the main W28 
shell.  Future observations should clarify if the nonthermal shell 
fragment is either part of W20 or yet another previously unidentified 
shell-type SNR.

\subsection{G6.83-0.21: A Shell-type Structure}

Lastly, we detect a weakly emitting shell-like source G6.83--0.21 
lying between M20 and W28, as labelled in Figure 6.  This source lies 
in a difficult region of our 20cm image where a ``negative bowl'' 
surrounding both W28 and M20 suppresses any emission arising from this 
weak source.  In spite of this difficulty, we note a shell source with 
a diameter of about 15$'$ lying between the northern and southern 
edges of W28 and G7.06--0.12, respectively.  The eastern edge of of 
G6.83--0.21 is brighter and has a typical flux density of about 4 mJy 
beam $^{-1}$ above the negative depression at a level of --6 mJy beam 
$^{-1}$.

It is difficult to determine the nature of G6.83--0.21 because of the 
lack of spectral and polarization information.  However, two 
observational anomalies, described below, imply that G6.83--0.21 is 
possibly a distinct shell-type SNR. The bright compact source 
6.833--0.093 and the pulsar PSR 1758-23 lie at the geometrical center 
of the shell source G6.83--0.21.  An association between PSR 1758-23 
and the W28 SNR has been suggested on the basis of age, even though 
the pulsar lies outside the W28 shell.  This hypothesis has the 
difficulty that it has a much larger dispersion measure than expected 
if the pulsar were placed 2kpc away (Kaspi et al.  1993), although 
Frail, Kulkarni and Vasisht (1991) have argued that the large 
dispersion and scattering of the pulses from PSR 1758-23 are caused by 
a dense screen of ionized gas located along the line of sight.  
Another anomaly is the kinematics of the OH(1720 MHz) masers observed 
across the northern part of W28 (Claussen et al.  1998).  OH(1720 MHz) 
masers have recently been identified as a signature of SNRs 
interacting with molecular clouds (e.g.  Frail, Goss \& Slysh 1994; 
Lockett, Gauthier \& Elitzur 1999; Wardle 1999).  The kinematics of 
these masers are generally found to be similar to the systemic 
velocity of molecular clouds into which SNRs are expanding.  Claussen 
et al.  (1998), however, notice an exception in W28, where the masers 
fall into low and high velocity groups.  Two molecular clouds have 
been observed toward W28 at 7 and 17 \kms\ in various molecular lines 
(Wooten 1981).  To explain this velocity difference, Claussen et al.  
(1998) suggest that the entire low-velocity molecular cloud has been 
accelerated by the SNR shock along our line of sight.

The above anomalies may be resolved if G6.83--0.21 is a shell-type SNR 
associated with the PSR 1758-23 and is interacting with a distinct 7 
\kms molecular cloud.  The distance to G6.83--0.21 is difficult to 
estimate but it has to be placed at a larger distance than 2 kpc to be 
consistent with the high dispersion of pulsar signals.  Future 
observations of G6.83--0.21 should be important to clarify if this 
feature is a confusing source in this complex region of the Galaxy or 
it is a new member of the class of SNRs with associated OH(1720 MHz) 
masers lying within the inner several degrees of the Galactic center 
(Yusef-Zadeh et al.\ 1999).

\acknowledgments

F. Yusef-Zadeh's work was supported in part by NASA.  
Basic research in radio
astronomy at the Naval Research Laboratory is supported by the Office of
Naval Research.



\newpage

\section*{Figure captions}

\figcaption{Contours of total intensity at 
$\lambda$3.6cm with a spatial resolution of 0.63$''\times0.45''$
(PA=80$^0$) with an rms noise of 33$\mu$Jy. The levels are set at 
-3, 3, 5, 7, 9, 11, 13, 15, 20, 25, 30, 40 times the rms noise.}

\figcaption{Exploratory NSFCAM images (log-scaled) of the central cluster of
M20.  The visible components of the multiple star HD 164492 are labelled in
the J image.  Field of each image is approximately 80$''\times80''$, with N up
and E left.  The dark spots are due to stars in the nearby blank sky
image which is used to subtract the background.}

\figcaption{A $\lambda$6cm radio  continuum 
image (3a) with a resolution of 2.3$''\times1.1''$ (PA=-8$^0$) 
is shown against  
a Palomar image from Digital
Sky survey (3b). Both images  show 
an  identical region 
and prominent features are identified in  both images. }

\begin{figure}
\plotone{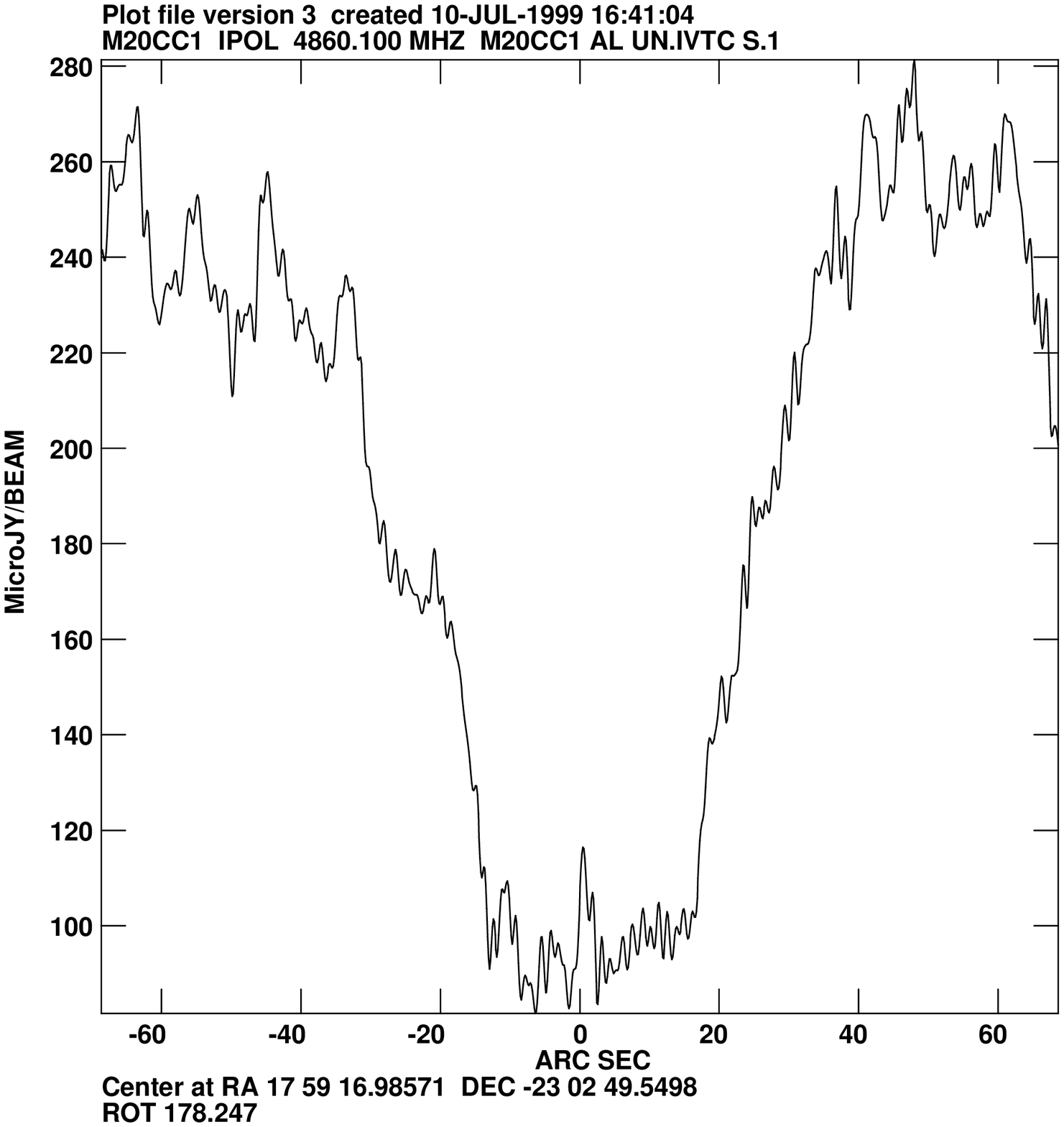} 
\figcaption{A slice cut along the EW dust lane
at 6cm centered at 
$\alpha, \delta(1950) = 17^h 59^m 17^s, -23^0 02' 49.5''$ 
at the position angle of 178$^0$.}
\end{figure}

\figcaption{A close up view of figure 3a showing 
TC1 and TC2 regions with contour levels set 
at -4, 4,, 6, 8, 10, 12, 14, 16 times 40 $\mu$Jy beam$^{-1}$.}

\figcaption{Greyscale image of the M20 region   
$\lambda$20cm with a spatial resolution of 65.2$''\times39.3''$
(PA=-34$^0$) identifying prominent sources.
The primary beam correction has not been applied to 
this image.} 

\figcaption{Contours of total intensity at
$\lambda$20cm with a spatial resolution of 65.2$''\times39.3''$
(PA=-34$^0$) with an rms noise of 1 mJy. The levels are set at
-3, 3, 5, 7, 9, 11, 13, 15, 20, 25, 30, 35, 40, 45, 50, 60, 70,
80, 100, 120, 140, 170, 200, 250, 300, 350 times 2 mJy beam$^{-1}$.} 

\figcaption{Contours of $\lambda$90cm emission obtained from
D-array configuration of the VLA. The angular resolution is 
5.5$'\times2.6'$(PA=13.4$^0$) 
and the rms noise is about 200 mJy beam$^{-1}$. 
The peak emission is 14.2 Jy beam$^{-1}$ and contour levels are 
at  100 mJy beam$^{-1}$ $\times$ -8, -6, 
-4, -2, -1, 1, 2, 4, 6, 8, 10, 20, 30, 40,
50, 60, 70, 80, 90, 100, 110, 120, 130.}

\begin{table}
\plotone{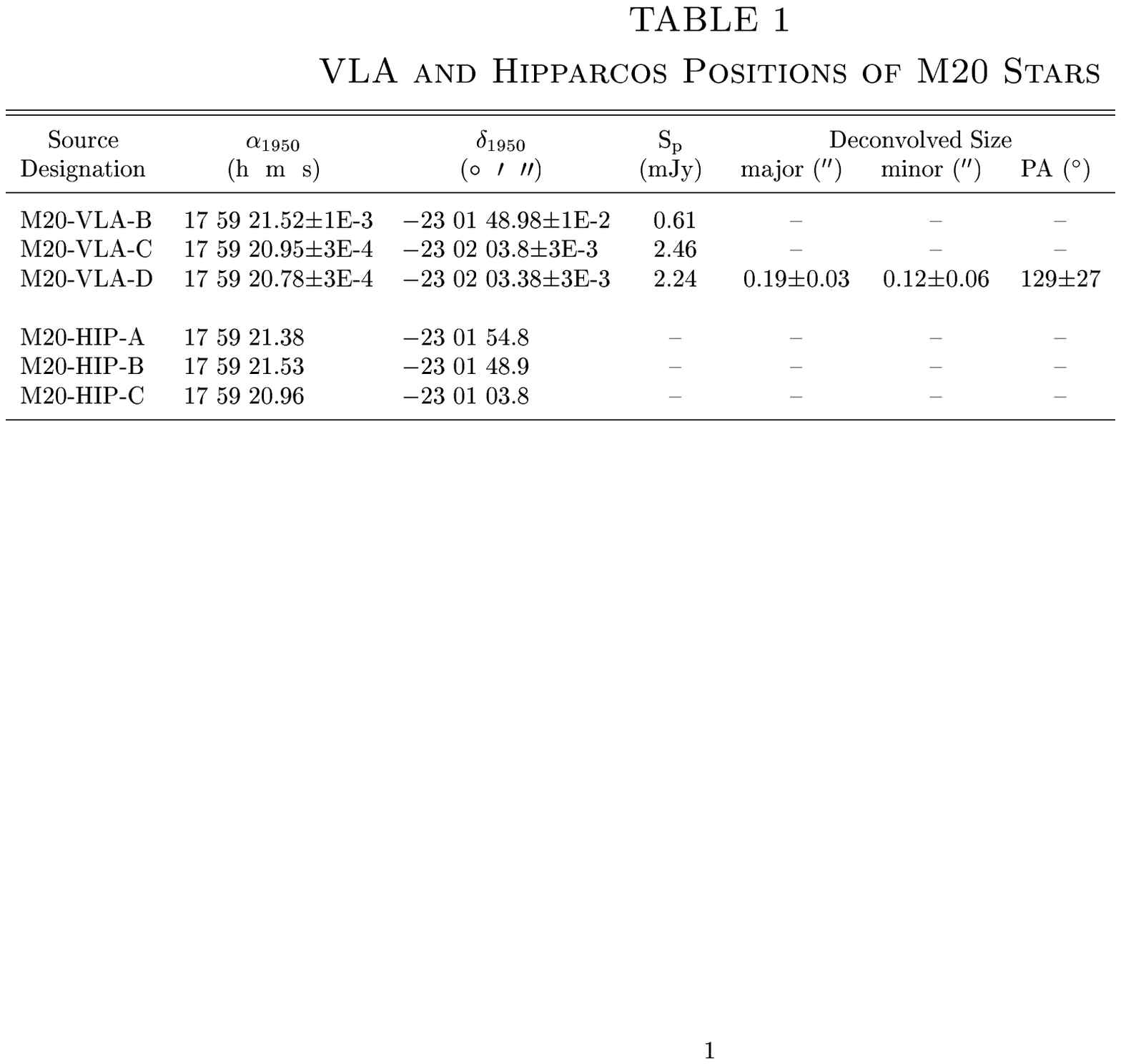}
\figcaption{Table 1}
\end{table}

\end{document}